\documentclass[aps,showpacs,twocolumn]{revtex4}
\usepackage{textcomp}
\usepackage{graphicx}
\begin{document}
\title{On the brink of jamming: Granular convection in densely filled
containers}

\author{Frank Rietz and Ralf Stannarius}

\affiliation{Otto--von--Guericke--University, D--39106 Magdeburg, Germany}

\date{\today}

\begin{abstract}
Granulates are ubiquitous in nature and technology, but despite their great importance, their
dynamics are by far less well understood than those of liquids. We demonstrate in an almost
compactly filled flat (Hele--Shaw) cell, where slow horizontal rotation simulates a variable
gravitational force, that unexpected dynamic structures may arise under geometrical restrictions.
The cell motion drives regular flow in the compact interior, and convection rolls combine with
segregation. The container fill level is crucial for the dynamic regime. A transition from chute
flow at lower fill levels to convection in densely packed containers is found. These observations
suggest the existence of comparable phenomena in situations where so far no systematic search for
dynamic patterns has been performed.
\end{abstract}

\maketitle

%\section{INTRODUCTION}
%\label{sec:introduction}

\begin{figure}[!h]
\begin{center}
\includegraphics[width=0.95\columnwidth]{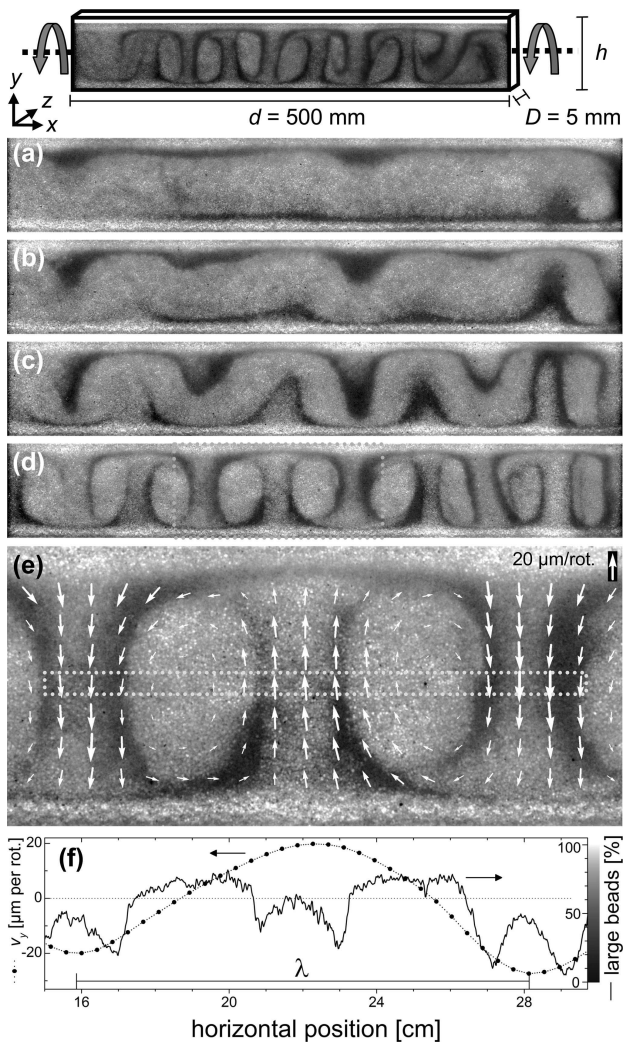}
\end{center}
\caption{
 Convection patterns in a densely filled cell. Top:~Sketch of the experimental geometry (`front view' of the cell),
 the illumination source is behind the cell, the camera is positioned in front.
 The free volume above the granulate represents a few percent of the cell volume.
 (a--d)~Development of the pattern after (a)~2,000, (b)~4,000,
 (c)~6,000 and (d)~12,000 rotations. Regions enriched with small particles
 appear darker, regions enriched with large particles appear brighter, the optical
 transmission reflects an average particle concentration through the cell depth.
 $C$\,=\,0.647\,$>$\,$C_{c}$, $h$\,=\,80\,mm. (e)~Detail of the texture of (d) and superimposed arrows
 of the flow field, i.e. the mean displacement of clusters per rotation, as detected from velocimetry (see text).
 (f)~Vertical flow velocity (dots) and composition of the mixture (solid line) in the central region of (e).
 The local composition has been determined from the optical transmission intensity.
 The wavelength\,$\lambda$ of the flow pattern is indicated.}
\label{fig1}
\end{figure}

The broad interest in the fundamental physics of granulates is motivated by their great relevance
in everyday life, in particular in the processing and handling of food, pharmaceuticals and
chemicals \cite{Knowlton:1994}. The complexity of granulates has so far withstood a description
within a unified theory. Phenomena related to dense flow and jamming have become central topics of
research activities \cite{Pouliquen_GDR,Richard_Trappe_Liu_Anna}. During the past decade, there has
also been an increased interest in pattern formation of granulates
\cite{Jaeger:1996,Kakalios_Aranson_Ottino}. One of the unique,
counterintuitive features of granular mixtures, for example, is their segregation under external
excitation. Among the classical systems are e.g. the horizontally rotating mixer \cite{Levine_Hill}
and horizontally or vertically vibrated layers \cite{Goetzendorfer:2006,Knight_Wassgren_Aoki,
Caballero:2004,Ristow:2000,Kudrolli:2004,Reis_Galanis}. In most experiments, the rearrangement of
individual grains in the container requires the fluidization of the granular bed, but slow fluxes
can occur even in deep layers where the grains do not have individual degrees of freedom for
positional changes. An impressive example is patterning of permafrost ground by freeze--thaw cycles
\cite{Kessler:2003}.

In densely packed systems, geometrical restrictions prevent the independent motion of particles,
and slow and collective dynamics are observed, as in the dynamic glass transition of supercooled
liquids \cite{Richard_Trappe_Liu_Anna}.
Very few studies consider such systems as e.g. densely filled
vibrating containers \cite{Caballero:2004} or horizontally rotating cylindrical mixers
\cite{Nakagawa_Turner,Kuo:2006}. Access to the inner structure of the granular bed is often
limited to invasive methods, and conclusions on subsurface convection have to be derived
indirectly. A direct hint to the bulk dynamics of such systems comes from numerical studies.
Simulations of disks in a flat, horizontally rotating cell \cite{Awazu:2000} revealed a global
convection flow. This motivated our experimental study of a flat, almost completely filled cell
rotating slowly about a horizontal axis. We observe unexpected regular convection structures in
combination with size segregation of the granulate (see Fig.~\ref{fig1} and Ref.~\cite{epaps}).

%\section{EXPERIMENT}
%\label{sec:experiment}

The measurements are performed in a transparent Hele--Shaw cell (Fig.~\ref{fig1}, top) with a
thickness of $D$\,=\,5\,mm, corresponding to a few particle diameters.
The cell length in axial direction is $d$\,=\,500\,mm, the height
$h$ is variable between 40\,mm and 110\,mm. The dry granulate consists of spherical glass beads
(W\"{u}rth Ballotini MGL) with density 2.46\,g/cm$^{3}$. Bimodal mixtures contain equal weight
fractions of small (250--350\,$\mu$m diameter) and large (800--900\,$\mu$m diameter) beads. The
cell is filled with the well mixed granulate, sealed, and thereafter rotated about the long
horizontal axis. The rotational speed is 20\,rpm. The Froude number, representing the ratio of
centrifugal and gravitational forces, is well below 0.025, thus centrifugal forces are negligible.
In the flat cell geometry, the principal effect of the rotation is a periodic modulation of the
effective gravitational acceleration, in combination with friction of the granulate at the two
plates of the Hele-Shaw cell. The system is in some respect similar to the vertically shaken
container \cite{Goetzendorfer:2006,Knight_Wassgren_Aoki,Caballero:2004,
Ristow:2000,Kudrolli:2004%,Galanis:2006
}. We note in passing that unlike in recently described convection rolls in a Taylor--Couette
experiment \cite{Conway:2004}, the granulate is $not$ slackened by air flow here.

First, it is necessary to define a quantitative measure for the fill level. The motional degrees of
freedom of the granular particles are controlled by the available space above the granulate.
However, this free volume is not a practicable parameter, since the height of the granular bed
depends on the filling procedure, and it is not exactly preserved during the experiment. Therefore
we weigh the granulate before filling. Using the mass and known density, we define the fill ratio
$C$ as the proportion of the net volume of glass beads to the cell volume. Note that, by this
definition, the free volume above the granulate for a given $C$ increases with the container height
$h$. For a cell completely filled with the bimodal mixture, without and with additional manual
tapping, we find $C$\,$\approx$\,0.66 and $C$\,$\approx$\,0.70, respectively.

The cell is observed in transmission with uniform background illumination. Images are automatically
recorded every 20 revolutions: rotation is paused when the cell is in vertical position and a
picture is taken with a CCD camera (KODAK Megaplus 6.3i, 3072 $\times$ 2048\,pixels).
A typical experiment runs for more than 10,000 rotations (one to a few days).

The collective particle motion is exploited for the determination of local cluster velocities.
Images are subdivided into evaluation squares of 8\,mm $\times$ 8\,mm and, within each segment, the
principal cluster displacement (maximum cross--correlation vector) is determined from the analysis
of consecutive images. This method allows a reconstruction of the local granulate motion in the
cell plane. Furthermore, the relation between the local transmission intensity and granular
composition has been established in separate experiments.

%\section{RESULTS}
%\label{sec:results}

\begin{figure}[htbp]
\begin{center}
\includegraphics[width=0.9\columnwidth]{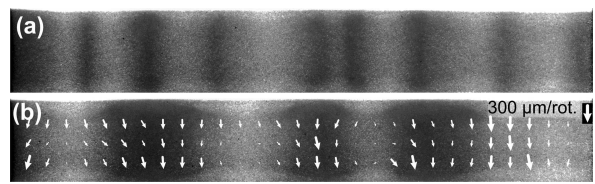}
\end{center}
\caption{
  Axially segregated stripe pattern in the chute flow regime,
  (a)~after formation of the segregation stripes, 1,500 rotations,
  (b)~coarsening after 15,000 rotations. Arrows symbolize the flow on the front side.
  $C$\,=\,0.581\,$<$\,$C_{c}$, $h$\,=\,80\,mm
  }
\label{fig2}
\end{figure}

The redistribution of particles during the cell rotation is basically restricted to the sliding of
the granulate, driven by gravity, when the cell plane tilts out of the horizontal. Two
qualitatively different regimes are found, depending on the fill ratio~$C$. Below some critical
$C_{c}$ (e.g. $C_{c}$\,$\approx$\,0.60 for the $h$\,=\,80\,mm cell), chute flow occurs twice in
each rotation cycle. Its result is the spontaneous formation of an axial segregation pattern
(Fig.~\ref{fig2}) of alternating stripes of small and large particles. A similar phenomenon is well
known from rotating cylindrical mixers \cite{Levine_Hill}. In the course of the experiment this
segregation pattern slowly coarsens, finally leaving two or three well segregated regions.

\begin{figure*}
\begin{center}
\includegraphics[width=\textwidth]{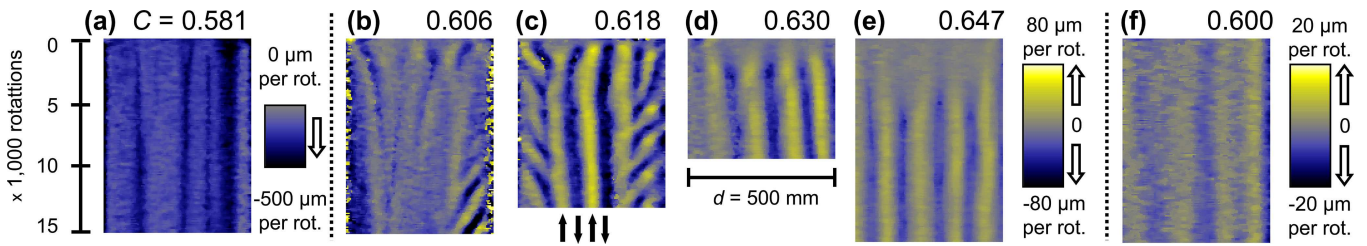}
\end{center}
\caption{(Color online)
 Time--space plots of the vertical flow field component $v_{y}$
 in the central axis of the cell. (a--e)~bidisperse mixture,
 $C_{c}$\,$\approx$\,0.60, and (f)~monodisperse granulate.
 Yellow (bright) indicates upflow, blue (dark) marks downflow. ($h$\,=\,80\,mm)
 (a)~global downward flow at a slightly subcritical $C$
 (see Fig.~\ref{fig2}), (b--e)~convective flow in dependence on the fill ratio~$C$.
 The flow field is uniform through the cell (in viewing direction),
 except for a very small global downstream component on the front side,
 of a few percent of the convection amplitude. When the cell is rotated by 180\textdegree{},
 up and down stream regions are interchanged. Images of Fig.~\ref{fig1}(a--d)
 correspond to the space--time plot\,(e).
 (f)~Flow field in a nearly monodisperse granulate.
 Note that monodisperse granulate has a lower packing density, the free volume above
 the granulate is comparable to that in\,(d).
  }
\label{fig3}
\end{figure*}

At $C_{c}$, the chute flow disappears and the granulate jams. At fill ratios above $C_{c}$, one
observes the formation of characteristic zones after the start of the experiment
[Fig.~\ref{fig1}(a)]. In shallow layers within a few mm of the top and bottom cell edges, the
particles' motion is fast and relatively uncorrelated. This region becomes enriched with large size
beads, indicated in Fig.~1(a--d) by the bright horizontal stripes. Except for these regions, the
grains can manoeuvre only in compact clusters (`glassy dynamics'). Within the next few mm, narrow
bands of jammed small size beads develop along each edge, while the central part of the cell
contains mixed granulate [Fig.~\ref{fig1}(a)]. A close inspection shows that dark regions contain
almost completely segregated small beads, while the bright regions resemble a mix dominated by the
large species. This initial structure becomes distorted by the onset of convective flow so that
small particles from the outer bands are advected towards the center [Fig.~\ref{fig1}(b)], and a
serpentine structure develops [Fig.~\ref{fig1}(c)]. Convection can start anywhere in the container.
The emerging rolls are decorated by small particles [Fig.~\ref{fig1}(d)]. The flow and composition
patterns seen on the front extend through the whole cell depth and there are only minor differences
between front and back textures. The described patterns and dynamics are qualitatively different
from those at low fill ratios. The periodicity of the convection pattern, once it has formed, is
persistent, and there is no coarsening. We note that different rotation speeds (between 5 and
20\,rpm) or evacuation of the interstitial air (between 9 and 101\,kPa) have no influence on this
general behavior. Similar convection patterns are observed for other materials too (e.g. glass bead
and poppy seed mixtures, sand mixtures).

For a quantitative analysis, we focus on the velocity field, determined from the displacement of
clusters as described above. Convective flow slowly sets in after the start of the experiment.
Figure~\ref{fig1}(e) shows exemplarily the flow field in a section of the cell. In the two
fluidized layers at the top and bottom edges, the flow field could not be determined. The particle
motion is rather uncorrelated there, and one to two orders of magnitude faster than in the
clusters. Figure~\ref{fig1}(f) shows the velocity $v_{y}$ (dots) and the local composition of the
mixture (solid line) obtained from the optical transmission intensity in the middle part of
Fig.~\ref{fig1}(e), vertically averaged between the dashed lines.

The vertical velocity $v_{y}(x,h/2)$ in the central cut is presented in Figs.\,\ref{fig3}(a--e) for
different fill ratios. Figure~\ref{fig3}(a) is a space--time plot of the flow field at subcritical
$C$\,=\,0.581 for an axially segregated stripe pattern. The global downward flow reflects a small
effective difference of the motion at front and rear planes, i.e. an effective vortex in the
$yz$\,plane, in the same sense as the cell rotation. Size segregation modulates the flow: in
general the regions of higher concentrations of small particles move faster. Figures
\ref{fig3}(b--e) evidence the qualitative change to the convection regime; the convection rolls are
manifest in spatially alternating up and down flows. Slightly above $C_{c}$, there is a strong
oscillation at the lateral edges of the cell [Fig.~\ref{fig3}(c)], possibly reflecting a
competition of rudimentary chute flow with the convection pattern. At higher fill ratios, these
oscillations vanish. The velocity amplitude decreases with increasing $C$, and the particle motion
becomes more and more collective. When $C$ approaches a value of about 0.70, no motion is
observable any more. Maximum velocities are of the order of 50\,$\mu$m/revolution. Thus, a cluster
needs several thousand cell rotations for a full turn. There is a small constant negative offset of
$v_y$, more than one order of magnitude smaller than the flow maxima.

From a Fourier analysis of $v_{y}(x,h/2)$, the mode selection is monitored. Boundary conditions
require that rolls end with up or down flow at the cell sides, we observe only half-integer and
integer modes. A selected wave number $q_{x}$\,=\,$d/\lambda$\,=\,4.5 in Fig.~3(e) corresponds to
4.5 roll pairs. In most cases, the fastest growing mode dominates from the beginning and determines
the wavelength of the pattern. In some experiments, the selected wavelength fluctuates between
neighboring modes. The spatial period of the flow structures is not sensitive to the fill ratio
above $C_{c}$. Figure~4 shows the influence of the container height on the selected wave number.
The dots label runs where the selected mode was constant, while vertical bars indicate that the
pattern showed temporal fluctuations of the selected mode during the experiment. As in typical
hydrodynamic convection structures (e.g. Rayleigh--B\'{e}nard) the selected wavelength is roughly equal
to twice the cell height. The relation $q_{x}$\,=\,$d/(2h)$ that would correspond to circular
convection rolls is indicated by a dashed line. The general trend is, in view of the small aspect
ratio of the experiment, in reasonable agreement with such an assumption.

\begin{figure}[htbp]
\begin{center}
\includegraphics[width=0.9\columnwidth]{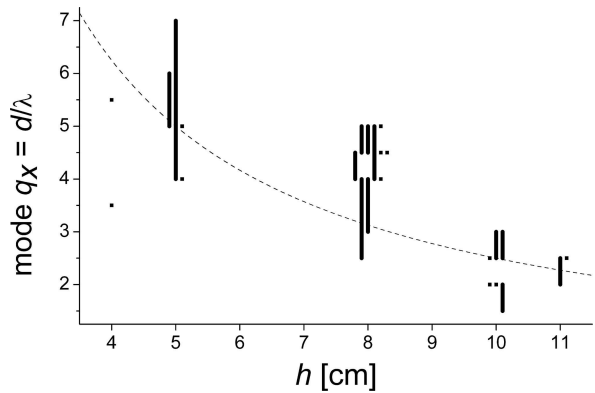}
\end{center}
\caption{
 Range of observed modes in dependence upon the cell height $h$. The data represent 24 individual experiments.
 The dashed curve symbolizes the relation for circular rolls,
 $q_{x}$\,=\,${d}/(2{h})$. For clarity, individual experiments for the same cell height
 are slightly displaced on the abscissa.
  }
\label{fig4}
\end{figure}

Finally, we have tested whether the convective flow is triggered by the segregation of the granular
mixture or whether the segregation of the mixture is a consequence of the convective flow. For that
purpose, a cell was filled with the nearly monodisperse beads of the large species ($<$\,10\,$\%$
radius variation). It is evident from Fig.~\ref{fig3}(f) that a convection structure of similar
wavelength as in the mixture is formed, but with considerably lower flow amplitude. Obviously the
bidispersity of the mixture strongly enforces the convection pattern, and the decoration by the
segregation structures facilitates the optical recognition of the pattern. However, the mechanism
works in a nearly monodisperse granulate as well, though much less effectively.

%\section{DISCUSSION}
%\label{sec:discussion}

The increasing collectivity of particle motion in clusters with increasing container fill ratio in
the rotating Hele--Shaw cell coincides with observations in vibrated beds \cite{Reis_Galanis}. We
observe the regular convection patterns in the high fill ratio range ($C$ between the upper bound
chute flow limit, $\approx$\,0.60, and the maximum achieved fill level, $\approx$\,0.70). This
convection looks morphologically similar to familiar structures in vibrated containers
\cite{Knight_Wassgren_Aoki,Caballero:2004,Ristow:2000,Kudrolli:2004}, but irrespective of this
superficial similarity there are striking differences. In the vibration experiments, up and down
accelerations are asymmetric. Friction at the side walls plays an essential role, and the rolls
appear generally pairwise \cite{Knight_Wassgren_Aoki,Caballero:2004,Ristow:2000,Kudrolli:2004}. A
critical acceleration is required \cite{Ristow:2000}, and the observed segregation patterns
\cite{Kudrolli:2004} are different. Similar to our experiment, convection is suppressed in the
vibrated container when the free volume above the granular bed is sufficiently reduced
\cite{Caballero:2004}.

Three mechanisms for granular convection have been discussed in literature
\cite{Jaeger:1996,Ristow:2000,Kudrolli:2004}, related to inhomogeneous agitation, friction at
lateral (left/right) container ends, and interactions with the interstitial fluid, respectively.
These are apparently irrelevant in our experiment. The slow dynamics of the granulate, and its practical independence of the revolution rate, suggest that dissipation
does not play a significant role in the pattern formation. Rather, local fluctuations of the
packing density could be relevant. Simulations \cite{Awazu:2000} for a 2D cell are in a certain
qualitative agreement with our experimental results, some arguments for a
convection mechanism are given, but they provide no direct clue on the nature of the instability
mechanism in our experiment.

Even though data for a truly three--dimensional system are not available, there are indirect
indications that convection structures in compactly filled containers are found in 3D systems too
\cite{Kuo:2006}. The experiments presented suggest the existence of slow subsurface currents in
similar situations that are not accessible to a direct optical analysis, and/or involve long,
geological \cite{Ortoleva:1994}, time scales.

%\begin{acknowledgments}
The authors are grateful to Tilo Finger, Nico Fricke, Jesse Gryn, Ulf Schaper, Stefan Scharfenberg,
Matthias Schr\"{o}ter and Christian Warnke for their advice and technical assistance. This study was
funded in part by a Landes\-stipendium Sachsen--Anhalt.
%\end{acknowledgments}

\end{document}